%% file: main.tex
\documentclass[preprint2,times,trackchanges]{aastex631}% \documentclass[preprint2,times,linenumbers,trackchanges]{aastex631}
\usepackage{graphicx}
\usepackage{amsmath}
\usepackage[figure,figure*]{hypcap}

\input{mycmd.tex}

\shorttitle{Sub-virial clumps as initial conditions}
\shortauthors{Wang, Wang, \& Xu}
\reportnum{Accepted to ApJL}

\begin{document}

\title{\large Massive Star Formation Starts in Sub-virial Dense Clumps Unless Resisted by Strong Magnetic Fields}
% \title{\large Most High-Mass Starless Clumps Are Collapsing Unless Resisted by\\ Strong Magnetic Fields}

\correspondingauthor{Ke Wang}
\email{kwang.astro@pku.edu.cn}

\author[0000-0002-7237-3856]{Ke Wang}
\affiliation{Kavli Institute for Astronomy and
Astrophysics, Peking University, 5 Yiheyuan Road, Haidian District, Beijing 100871, China}

\author{Yueluo Wang}
\affiliation{Department of Astronomy, School of Physics, Peking University, Beijing, 100871, People’s Republic of China}

\author[0000-0001-5950-1932]{Fengwei Xu}
\affiliation{Kavli Institute for Astronomy and
Astrophysics, Peking University, 5 Yiheyuan Road, Haidian District, Beijing 100871, China}
\affiliation{Department of Astronomy, School of Physics, Peking University, Beijing, 100871, People’s Republic of China}
\affiliation{I. Physikalisches Institut, Universität zu Köln
Zülpicherstr. 77, 50937 Köln, Germany}

\begin{abstract}

The initial conditions 
are critical for understanding high-mass star formation, but are not well observed. 
Built on our previous characterization of a Galaxy-wide sample of 463 {candidate} high-mass starless clumps (HMSCs), here we investigate the dynamical state of a representative subsample of 44 HMSCs {(radii 0.13-1.12\,pc)} %median 0.56 pc 
using GBT \nh3\ (1,1) and (2,2) data from the Radio Ammonia Mid-Plane Survey (RAMPS) pilot data release.
By fitting the two \nh3\ lines simultaneously, we obtain velocity dispersion, gas kinetic temperature, \nh3\ column density and abundance, Mach number, and virial parameter.
Thermodynamic analysis reveals that most HMSCs have Mach number $<$5, inconsistent to what have been considered in theoretical models.
All but one (43/44) of the HMSCs are gravitationally bound with virial parameter $\alpha_{\mathrm{vir}} < 2$. 
Either these massive clumps are in collapsing or magnetic field strengths of 0.10-2.65\,mG (average 0.51\,mG) would be needed to support them against collapse. 
The estimated B-field strength correlates tightly with density,
$B_{\rm est}/{\rm mG}=0.269\,(n_{\rm H_2}/10^4\,{\rm cm^{-3}})^{0.61}$,
with a similar power-law index as found in observations, but a factor of 4.6 higher in strength.
For the first time, the initial dynamical state of high-mass formation regions has been statistically constrained to be sub-virial, in contradictory to theoretical models in virial equilibrium, {and in agreement with the lack of observed massive starless cores}.
The findings urge future observations to quantify the magnetic field support in the prestellar stage of massive clumps, which are rarely explored so far, towards a full understanding of the physical conditions that initiate massive star formation.
% The Five-hundred-meter Aperture Spherical radio Telescope (FAST) 
% FAST can make a significant contribution with demonstrated Zeeman splitting feasibility.

\end{abstract}


% \keywords{Star formation (1569)
% --- Star forming regions (1565)
% --- Infrared dark clouds (787)
% --- Interstellar magnetic fields (845)
% --- Interstellar medium (847)
% }

% https://authortools.aas.org/cgi-bin/newlatexwordcount.cgi
% https://astrothesaurus.org/thesaurus/alphabetical-browse/
% from within AAS submission, select keywords:
% Star formation (1569), Star forming regions (1565), Interstellar filaments (842), Interstellar magnetic fields (845), Interstellar medium (847), Galaxy structure (622), Catalogs (205)

% Star formation (1569), Star forming regions (1565), Infrared dark clouds (787), Interstellar magnetic fields (845), Interstellar medium (847), Molecular clouds (1072), Collapsing clouds (267)

\section{Introduction} \label{sec:intro}

Massive stars ($M_\star \gtrsim8$\msun) play an irreplaceable role in the energy budget and evolution of the Galactic ecosystem. However, their origin remains a fundamental open question. Particularly, the initial conditions are not well constrained by observations, leading to different assumed initial inputs to theoretical models. For example, in the long-standing turbulent core model {\citep{McKee_Tan2002Natur,Mckee03-formmodel}}, massive stars form within massive ($\sim$100\msun) pre-assembled, monolithic (non-fragmenting) cores in highly supersonic turbulence with Mach number $\gtrsim$5, proceeding in virial equilibrium such that the cores do not collapse free-fall. 
{The model is built on self-similar cores and clumps, where the cores are embedded in clumps, and both share similar dynamical properties \citep{Bertoldi92-virial, McKee_Tan2002Natur,Mckee03-formmodel}}.
In contrast to this ``core-fed'' scenario, 
the competitive accretion \citep{Bonnell01-formmodel}, global hierarchical collapse \citep{Vazquez97-formmodel}, and inertial flow models \citep{Padoan20-formmodel} can be categorized as the ``clump-fed'' scenario, characterized by gas assembly through either global clump infall and/or coherent and filamentary gas flows \citep[e.g.][]{Peretto2013,Xu2023SDC335} that results in massive cluster formation.
{Both theoretical scenarios require highly turbulent gas; in particular the turbulent core accretion model requires Mach number \mach$\gtrsim$5 \citep{Mckee03-formmodel,Krumholz2007a,Krumholz2007b}; similar and even higher Mach numbers are adopted in numerical simulations focusing on turbulence regulated star formation, at clump and cloud scales \citep[e.g.,][]{Federrath2012}.
The turbulent core model assume virial equilibrium in self-similar cores and clumps as an initial condition, but do not assume so for protostellar and later evolutionary stages (this has been misinterpreted in some literature).}
The competitive accretion model requires \avir$<$1 {for a dynamical, non-equilibrium star formation.}
% Without considering magnetic fields, the turbulent core model requires virial parameter \avir$\gtrsim$2, while the competitive accretion model requires \avir$<$1.
It is therefore of critical importance to observe the initial conditions at the pre-stellar stage, before protostars have formed with strong feedback to drastically change the conditions.

Observationally, the early evolution of massive clumps is particularly uncertain, since cold, quiescent clumps are fainter at (sub)millimeter wavelengths and spectral lines than those hosting protostars or even HII regions. 
% However, they are essential to testing the theoretical models mentioned above since they are less polluted by the strong stellar feedback or dynamical evolutionary effects. 
For statistically significant studies, an unbiased sample of high-mass starless clumps (HMSCs) is necessary. Several studies have made significant efforts to identify reliable HMSCs \citep[e.g.,][]{Tackenberg2012,Traficante15-LM-rela,Svoboda2016,YuanJH17-HMSCs}.
Among these, we have presented a sample of 463 HMSCs \citep{YuanJH17-HMSCs} across the inner Galactic plane ($|l| < 60^\circ$, $|b| < 1^\circ$), combining ATLASGAL, HiGAL, {GLIMPSE, MIPSGAL, and literature data} \citep{Benjamin2003,Carey2009,Schuller09-ATLASGAL,Csengeri2014_ATLASGAL_cat,survey:HiGAL-DR1-Molinari2016}. 
% carefully identified from the 10861 ATLASGAL clumps 
HMSCs are massive and dense sufficient to form high-mass stars, but have not shown signatures of star formation.
{\cite{YuanJH17-HMSCs} make source selection through a strict and comprehensive work flow (see their Fig 1). Briefly, we started from all the 10861 ATLASGAL clumps, selecting a flux limited sample (peak flux $>$0.5 Jy/beam at 870\um) to ensure high-mass, removing clumps with star formation signatures using HiGAL and by querying the SIMBAD database. The signatures include YSOs, IR point sources at 8\um, 24\um, 70\um, extended 70\um\ emission, (candidate) outflows, masers, ``extended green objects'', HII regions, and radio continuum sources recorded in the SIMBAD database (see a full list of 18 signatures in their Table 1). 
Physical properties of the HMSCs are then derived by pixel-to-pixel SED fitting to combined HiGAL and ATLASGAL data. The 463 clumps satisfying these strict criteria form a complete unbiased sample of HMSCs across the Milky Way disk. Stickily speaking, they should be regarded as candidate HMSCs, because deeply embedded star formation activity may exist \citep[e.g.,][]{Feng2016,Cyganowski2022,JiaoWY23-HMSCs}, and may be under the detection limit of these single-dish shallow surveys. High-resolution interferometric observations are being carried out for this purpose \citep[e.g.][]{JiaoWY23-HMSCs}. Nevertheless, these HMSCs represent the earliest observed stages that defines the initial conditions for high-mass star formation.}

{The 463 HMSCs have a median radii of 0.65\,pc and range in 0.05-3.57\,pc. For convenience, we refer them as ``clumps'' throughout this paper, but note that they have a fairly wide range; the smallest of them (20 HMSCs) are actually more appropriate to be called as ``cores'' \citep{YuanJH17-HMSCs}.}

{We emphasize that the source selection criteria in \cite{YuanJH17-HMSCs} are the most strict ever employed in studies towards the initial stages. More recent characterization of HiGAL clumps, for example \cite{Merello2019}, classify prestellar clumps as those without 70\um\ point source. Other potential star formation signatures (e.g., recorded in SIMBAD literature) are not used. \cite{Merello2019} compiled a sample of 1068 HiGAL clumps which have single-pointed \nh3\ observations from the literature within half of the corresponding \nh3\ beam (mostly from \citealt{Wienen2012}, see next paragraph). Among those, only 157 are classified as prestellar by the authors, and at most only 19 would satisfy our selection criteria of HMSCs in \cite{YuanJH17-HMSCs}. 
% \cite{Merello2019} also caution that ...
\cite{Traficante18-sigma-R-rela} compiled a sample of HiGAL 213 clumps with \n2hp\ linewidth from the MALT90 survey, where only 14 clumps are starless. 
}

{\cite{Wienen2012} carried out single-pointed \nh3\ line observations towards a flux limited (peak APEX 870\um\ flux $>$0.4 Jy/beam) sample of 862 ATLASGAL clumps in the northern sky using the Effeleberg 100\,m telescope, with a beam of 40$''$ and spectral resolutions of 0.5 and 0.7\kms. Although the flux limit is higher than that of \cite{YuanJH17-HMSCs}, i.e., the clumps are all considered by our selection, only 43 among the 862 clumps ($<$5\%) fulfill our criteria for HMSCs \citep[as noted in][]{YuanJH17-HMSCs}. Additionally, the sparse spectral resolutions would poorly sample relatively narrow linewidth. \cite{Wienen2018} carried out similar \nh3\ observations towards 354 ATLASGAL clumps with peak 870\um\ flux $>$1.2\,Jy/beam in the southern sky using the Parkes 64\,m telescope, with a beam of 61$''$ and spectral resolutions of 0.4\kms. The beamsize is larger than all the clump sizes, making contamination a potential problem.}

{Thus, although $\sim$1000 unique (massive and low-mass) clumps have been studied by single-pointed \nh3\ and \n2hp\ lines, an unusually tiny fraction ($<$5\%) of those are at the starless stage. This is a result of complicated selection bias \citep{Merello2019}.
Due to limited observation time, \cite{Wienen2012,Wienen2018} did not complete ATLASGAL sources satisfying their selection criteria; due to sensitivity limit, not all observed sources are detected. The combined selection effect is difficult to characterize. Selection bias in the HiGAL follow-up works are even more difficult to quantify, because the \nh3\ observations are not dedicated follow-ups, and thus suffered from different sky coverage, source selection, positional offset, resolution, and completeness, among other issues, as noted in \cite{Merello2019,Traficante18-sigma-R-rela}. 
Because of these biases, it is highly questionable whether this tiny fraction of starless clumps is representative of the entire population of starless clumps. Consequently, it is extremely difficult, if possible at all, to identify the initial conditions from these existing results.}

{The tiny fraction of genuine starless clumps and the associated biases in the aforementioned ATLASGAL and HiGAL follow-up studies reinforces the necessity for a systematic characterization of the dedicated HMSC sample to reveal the initial conditions that would initiate high-mass star formation.
Properties of clumps at protostellar and later evolutionary stages presented in these studies provide a highly complementary comparison to investigate how physical conditions may change before and after star formation started.
}

\begin{figure*}[ht!]
\centering
\includegraphics[width=0.9\linewidth]{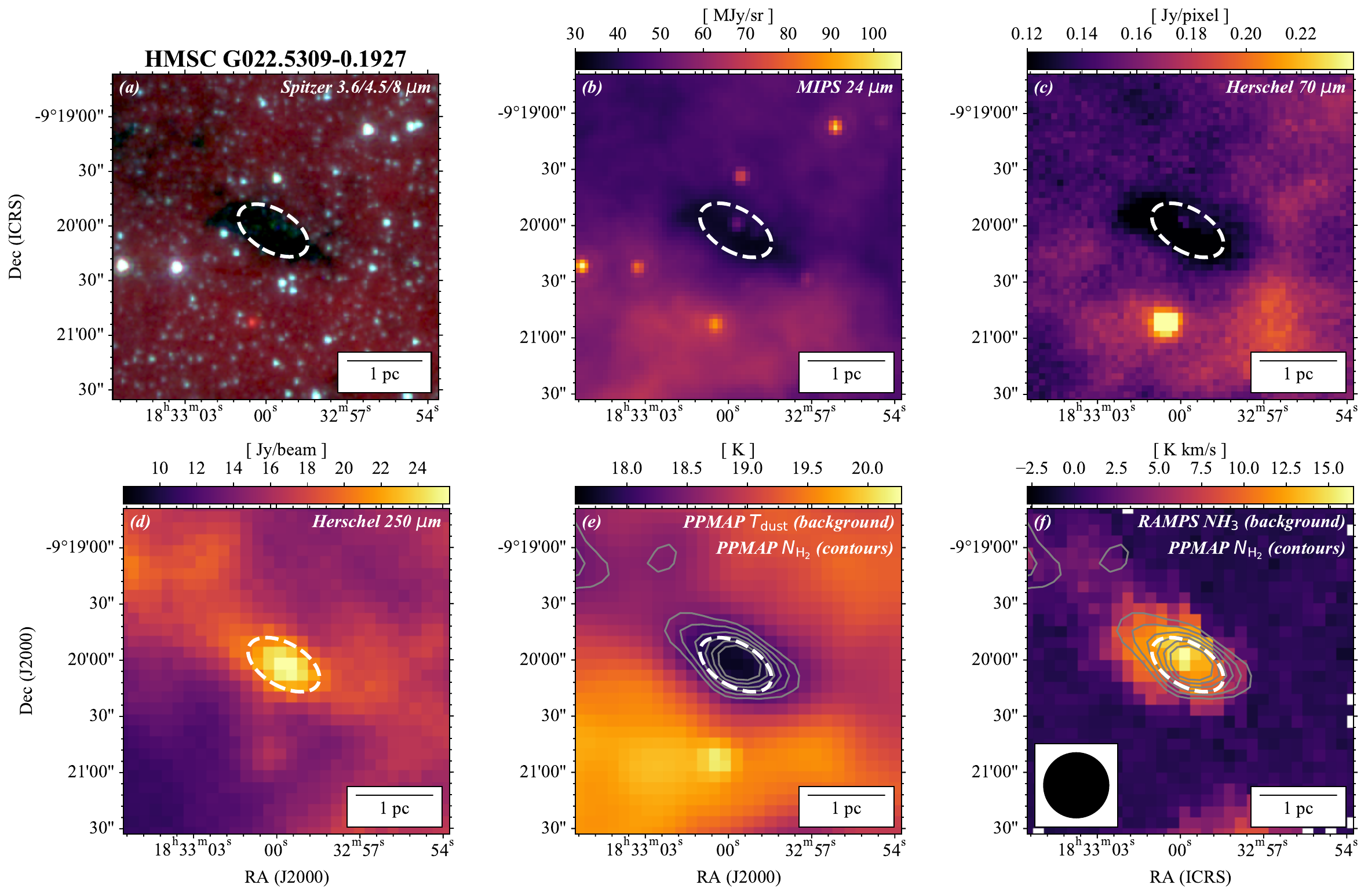}
\caption{Overview of an example high-mass starless clump (HMSC) G022.5309-0.1927.
% environments with multiwavelength maps and the RAMPS \nh3\ moment-0 map. 
(a) Three pseudo-color images with \textit{Spitzer} {maps} at 8.0, 4.5, and 3.6 $\mu$m {\citep{Benjamin2003}} rendered in red, green, and blue, respectively. (b) \textit{Spitzer} 24 $\mu$m emission {\citep{Carey2009}}. (c) \textit{Herschel} 70 $\mu$m emission {\citep{Molinari2010,Molinari2016}}. (d) \textit{Herschel} 250 $\mu$m emission. (e) the dust temperature is shown in background and H$_2$ column density in gray contours of [2.0, 2.5, 3.0, 3.5, 4.0, 4.5, 5.0]$ \times 10^{22}$ cm$^{-2}$ \citep{Marsh17_PPMAP}. (f) moment 0 map of $\mathrm{NH_{3}}$(1,1) data of RAMPS show a good spatial correlation with cold dust emission; the GBT beam size is shown at the bottom left corner. In all panels, the white dashed ellipse outlines the source size.}
\label{fig:rgb}
\end{figure*}

{In this context, we} studied the dynamical state of these HMSCs \citep{HuangB23-HMSC-CO}, using \cob\ spectra from the SEDIGISM survey \citep{Schuller2017_SEDIGISM,Schuller2021_SEDIGISM,Duarte2021_SEDIGISM} to obtain linewidth, finding that about half of HMSCs are gravitationally bound. 
% $\alpha\lesssim2$.
However, CO can be depleted in cold and dense environments in prestellar stages \citep{Goldsmith2001_depletion,Bergin2007ARAA_coldDarkClouds,ChenHR2010_G28_deu_dep,me2012}. Also, the optical depth of the \cob\ line is relatively high. 
% The systematic characterization of \cite{HuangB23-HMSC-CO} suffered from overestimation of kinetic energy using $^{13}$CO\,(3--2) linewidth.
These will lead to a larger linewidth and thus overestimate the kinetic energy. In fact, \cite{HuangB23-HMSC-CO} already noted that \nh3\ linewidth is systematically lower than that of \cob.
% On the other hand, 
\nh3\ is a reliable tracer for cold and dense gas without depletion, for example in infrared dark clouds \citep[IRDCs,][]{Pillai2006,qz2011,WangY2008,me2012,WangK14-HMSCs,WangChao2024_15irdcs,WangChao2023_G35.2}.
% (pillai, me, qz G30.88, Xie JJ, Chira, WangChao), 
\autoref{fig:rgb} shows an example HMSC. The \nh3\ emission traces the cold and dense clump very well.
Moreover, in addition to linewidth, \nh3\ lines provide a perfect thermometer, making it excellent to study thermodynamics in HMSCs \citep{Ho1983,Walmsley1983}.

In this work, we use \nh3\ (1,1) and (2,2) to trace dynamical state and temperature of the $\sim$pc scale HMSCs from \cite{YuanJH17-HMSCs}, an unbiased, well characterized sample for the initial stages of high-mass star formation.

\section{\nh3\ Data and Fitting} \label{sec:data-method}

The Radio Ammonia Mid-Plane Survey \citep[RAMPS,][]{Hogge18-RAMPS} {is an ongoing mapping survey} aims to map a portion of the first Galactic quadrant ($l=10^\circ$--$40^\circ$, $|b|<0.4^\circ$) using the Green Bank Telescope (GBT). We retrieved \nh3\ (1,1) and (2,2) spectral cubes from the RAMPS pilot data release\footnote{{\url{http://sites.bu.edu/ramps}}}, which have mapped about 6.5 square degrees in a series of ten fields centered at $l = 10^\circ, 23^\circ, 24^\circ, 28^\circ, 29^\circ, 30^\circ, 31^\circ, 38^\circ, 45^\circ$ and $47^\circ$ \citep{Hogge18-RAMPS}. 
% The GBT beam is $32''$ for the data, 
The \nh3\ spectral images have a FWHM beam of about $32''$, and the rms noise is about 0.16 K at a spectral resolution of 0.2\kms.
% , after channel smoothing.
\autoref{fig:rgb} shows an overview of the \nh3\ emission compared to multi-wavelength images of a typical HMSC.

{The RAMPS pilot surveyed area covered 51 of the 463 HMSCs} distributed in the inner Galactic plane \citep{YuanJH17-HMSCs}.
% We compared the range of the RAMPS data coverage with the identified HMSCs reported by \cite{YuanJH17-HMSCs} and find 51 valid sources 
% (6 clumps from $l = 10^\circ$, 18 clumps from $l = 23^\circ - 24^\circ$, and 20 clumps from $l = 28^\circ - 31^\circ$, respectively). 
We extracted the mean spectra of each of the HMSCs within the elliptical sizes provided by \cite{YuanJH17-HMSCs}.
{All the 51 HMSCs are detected in \nh3, demonstrating the robustness of the source selection and high quality of the RAMPS data.}

\begin{figure*}[hp!]
    \centering
    \includegraphics[width=0.819\linewidth]{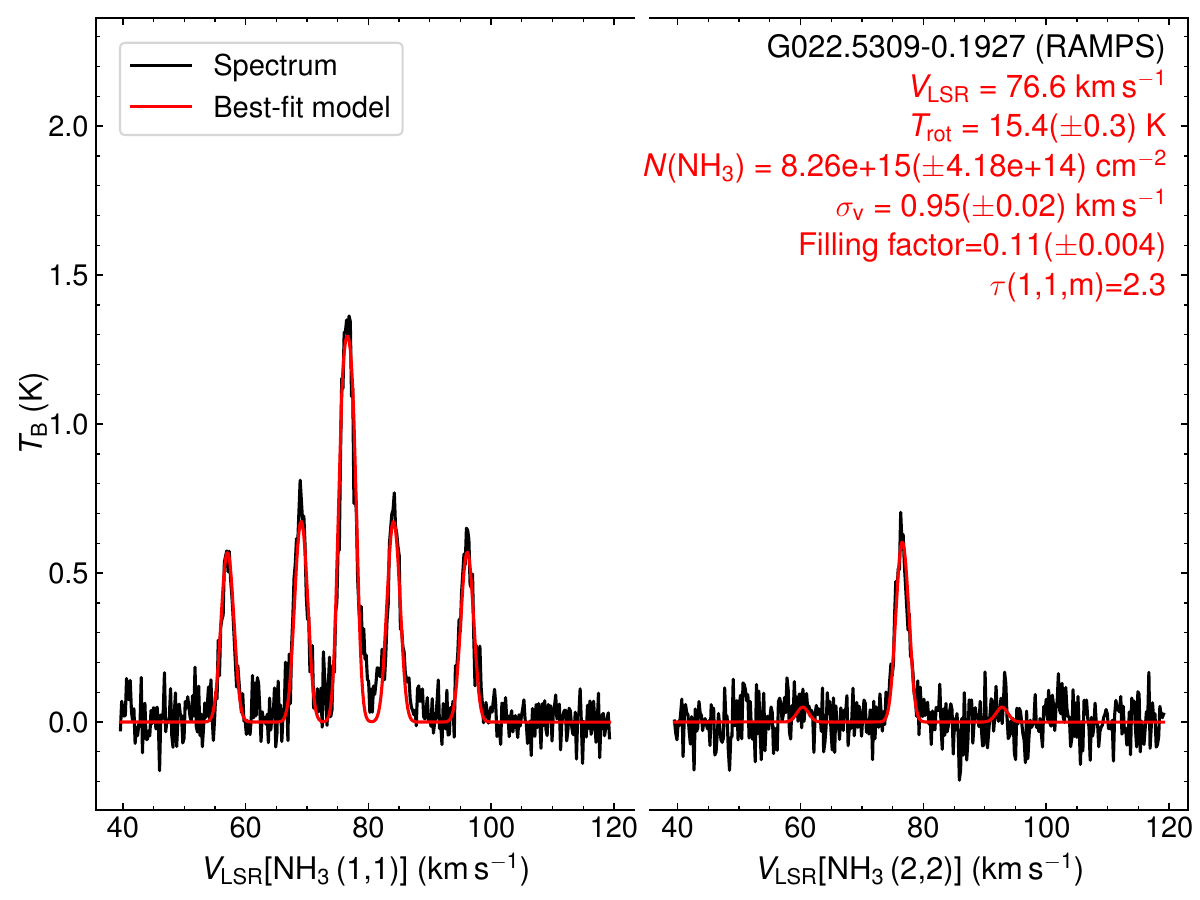}
    \caption{An example of the NH$_{3}$ line fitting using our procedure. The black lines are mean NH$_{3}$\,(1,1) and NH$_{3}$\,(2,2) spectra extracted from within the clump size of HMSC G022.5309-0.1927. The red lines show the best fitted model. The fitted parameters and their uncertainty are printed on the up-right corner of the plot.}
    \label{fig:spec}
\end{figure*}

We fitted the \nh3\ (1,1) and (2,2) spectra simultaneously using Pysepckit \citep{Ginsburg22-Pysepckit}, in a similar way as in \citep{WangK14-HMSCs,Sokolov18-velo-disper,WangChao2023_G35.2}. 
In the process of building the fitting program, we referenced some  codes from \cite{Lu14-Pyfit}, which uses a modular architecture to facilitate single-component and two-component fitting. 
% allowing users to verify the fitting results and analyze various properties. 
The procedure models the \nh3\ spectra with five free parameters:
centroid velocity $V_{\mathrm{LSR}}$, rotation temperature $T_{\mathrm{rot}}$, NH$_{3}$ column density $N_{\mathrm{NH_{3}}}$, line width in dispersion $\sigma$, and optical depth $\tau$.
We also added beam filling factor as a free parameter.
Among the 51 clumps, 7 show more than one velocity component in the spectra. We have removed them from further analysis, and focus on the 44 HMSCs with a single velocity component in this work.
{The 44 HMSCs represent properties of the full sample of 463 HMSCs.}
\autoref{fig:spec} presents an example for the fitting. 
% The fitted parameters are used for thermodynamic analysis below.

\section{Thermodynamic Analysis} \label{sec:analy}
% \subsection{Dynamical state of HMSCs}
\subsection{Derivation of thermal and dynamical parameters}

The \nh3\ fitted parameters are used together with clump mass and size from \cite{YuanJH17-HMSCs} for thermodynamic analysis of the HMSCs. 

The rotational temperature \Trot\ governs the level population of the \nh3\ system; it is related to kinetic temperature with collision coefficients. We use the empirical relationship\footnote{
{\cite{soft:Estalella2017_HfS} present a slightly modified relationship. For the minimum, median, and maximum \Tkin\ in Table 1, the modified version differs from \autoref{eq:Tkin} by 0.7\%, 1.8\%, and 4.5\%, respectively.}
% Compared to \autoref{eq:Tkin}, for the minimum, median, and maximum \Tkin\ in Table 1, the differences are 0.7\%, 1.8\%, and 4.5\%, respectively, and are therefore negligible for this work.
% it results to 0.3\,K difference in \Tkin\ for the median \Trot. The difference is negligible, and because this is only an empirical relationship, we keep using \autoref{eq:Tkin} in this work.
} deduced in \cite{Walmsley1983,Tafalla2004} to derive \Tkin\ from \Trot:
\begin{equation}\label{eq:Tkin}
    T_{\mathrm{rot}} = \frac{T_{\mathrm{kin}}}{1+\frac{T_{\mathrm{kin}}}{41.5}\,{\mathrm{ln}}[1+0.6\,{\mathrm {exp}}(-\frac{15.7}{T_{\mathrm{kin}}})]}
\end{equation}

% The level population is governed by the rotational temperature of the NH3 system, which is related to kinetic temperature with collision coefficients (Danby et al. 1988).

The line velocity dispersion is broadened due to the spectral resolution of $\Delta v_{\mathrm{chan}}=0.2$\kms. We subtract this minor broadening in quadrature
% (e.g., Longmore 07,11): Jijina

\begin{equation}\label{eq:sigV}
    \sigma_{\mathrm{v}}^{2} = \sigma^{2} - \frac{\Delta v_{\mathrm{r}}^{2}}{8\mathrm{ln}2}
\end{equation}
% as the sigV reported in Table 1.

For comparison with literature, FWHM linewidth is used more frequently than velocity dispersion. For a Gaussian profile as modeled in our work and widely used in literature, one have 
\begin{equation}\label{eq:dvFWHM}
\begin{aligned}
\Delta V_{\mathrm{FWHM}} 
& = 2\sqrt{2\mathrm{ln}2} \times \sigma_{\mathrm{v}} \\ 
& = 2.355 \sigma_{\mathrm{v}}
\end{aligned}
\end{equation}
% 2.354820045

With the gas temperature determined, the thermal broadening to velocity dispersion is

\begin{equation}\label{eq:sig_th}
\sigma_{\mathrm{th}}=
\sqrt{
\frac{k_{\mathrm{B}}T_{\mathrm{kin}}}{m_{\mathrm{NH_3}}}
}
\end{equation}
where
$k_{\mathrm{B}}$ is the Boltzmann constant, 
$T_{\mathrm{kin}}$ is the kinetic temperature, 
$m_{\mathrm{NH_3}}=17m_{\mathrm{H}}$ is the mass of the \nh3\ molecule,
$m_{\mathrm{H}}$ is the mass of the hydrogen atom.

The velocity dispersion due to non-thermal gas motion is then computed by subtracting that of the thermal motion in quadrature \citep{Myers83-velo-disper,Fuller92-velo-disper}
% (\citealp{Myers83-velo-disper}; \citealp{Fuller92-velo-disper}; \citealp{Sanchez13-velo-disper}; \citealp{Palau15-velo-disper}; \citealp{Henshaw16-velo-disper}; \citealp{Sokolov18-velo-disper})
\begin{equation}\label{eq:sig_nth}
  \sigma_{\mathrm{nth}} = \sqrt{
  \sigma_{\mathrm{v}}^{2}-\sigma_{\mathrm{th}}^{2}
  }
\end{equation}

The sonic Mach number, defined as the non-thermal velocity dispersion with respect to sound speed $c_s$, is then calculated as
% of the molecular hydrogen gas, which is referred to as the Mach number $\mathcal{M}_{S}$ 
(e.g., \citealp{Palau15-velo-disper})
\begin{equation}\label{eq:mach}
\begin{aligned}
\mathcal{M}_{S} 
&= \frac{\sigma_{\mathrm{nth}}}{c_{s}} \\
& = \frac{\sigma_{\mathrm{nth}}}{ \sqrt{(k_{\mathrm{B}}T_{\mathrm{kin}})/(\mu_{\mathrm{m}}m_{\mathrm{H}})} }    
\end{aligned}
\end{equation}

% where $c_{s}=\sqrt{(k_{\mathrm{B}}T_{\mathrm{kin}})/(\mu_{\mathrm{m}}m_{\mathrm{H}})}$ is the sound speed of the gas and 
where $\mu=2.33$ is the mean molecular weight.

Then we calculate virial mass and virial parameter to assess the dynamical state of the clumps, following \cite{Maclaren88-virial,Bertoldi92-virial}:
% For a clump without external pressure and magnetic fields, the virial mass is derived by \cite{Maclaren88-virial}. Consider a clump with density profile $\rho \propto r^{-n}$, the virial mass is expressed as
\begin{equation}\label{eq:Mvir}
\begin{aligned}
M_{\mathrm{vir}} 
&= 3k_1 %\left( \frac{5-2n}{3-n} \right)
  \frac{R\sigma^{2}_{\mathrm{v}}}{G} \\
&= 126k_1 %\left( \frac{5-2n}{3-n} \right)
\left(\frac{R}{\mathrm{pc}}\right)
\left(\frac{\Delta V_{\mathrm{FWHM}}}{\mathrm{km}\cdot \mathrm{s}^{-1}}\right)^{2}
M_{\odot}
\end{aligned}
\end{equation}
where $k_1 = (5-2n)/(3-n)$ is a correction factor to account for  nonuniform density distribution, which is characterized by a power law profile $\rho \propto r^{-n}$.
% where $n$ is the power law index in the density profile, 
$R$ is the equivalent radius of the clump,
and $G$ is the gravitational constant.
According to observed density profiles in high-mass star formation clumps, we adopt $n=1.8$ following \cite{Sanhueza2017},
% {Sanhuza, Lin, refs in.}, 
leading to $k_1 = 1.667$.

The virial parameter, defined as the ratio of the kinetic energy and gravitational energy, is then computed as 
% \citep{Bertoldi92-virial}
\begin{equation}\label{eq:a_vir}
  \alpha_{\mathrm{vir}} = k_2 \frac{M_{\mathrm{vir}}}{M_{\rm cl}}
\end{equation}
% where $M_{\mathrm{cl}}$ 
where $M_{\rm cl}$ is the clump mass, and $k_2$ is a correction factor to account for the clump's deviation from a sphere. \cite{Bertoldi92-virial} derived that $k_2$ depends only on the clump's aspect ratio, and for the ellipcity of the HMSCs considered here \citep{YuanJH17-HMSCs}, $k_2$ equals to unity within a few percent (see Eq. A8 and Fig 2 in \citealt{Bertoldi92-virial}; cf. \citealt{Henshaw16-velo-disper}).

The uncertainty in \avir\ mainly originates from uncertainties in source kinematic distance, dust emissivity, gas-to-mass ratio, \nh3\ linewidth and kinetic temperature (\autoref{fig:spec}). Following a comprehensive uncertainty evaluation in \cite{Sanhueza2017}, we quote a typical 75\% uncertainty in \avir.

The above derivations have neglected external pressure. Since the HMSCs are embedded within larger clouds, external pressure help to confine the clumps, and in term contritube to virial equilibrium. It has been shown that the surface energy term is of the same order as volume energy term \citep{Ballesteros06-surf-ene,Dib07-surf-ene}. Therefore, we follow \cite{Kauffmann13-Avir} and treat $\alpha_{\mathrm{vir}} < 2$ as gravitationally bound, instead of $\alpha_{\mathrm{vir}} < 1$ as the criterion.

\subsection{Estimated magnetic fields}

The above calculations have not considered magnetic fields, which is commonly observed in star formation regions \citep{LiuJH2022_Pol288,Pattle2023-PP7}.
% (ref, HullQZ, Junhao, Pattel). 
Magnetic fields provide additional support against gravity. The Alfvénic turbulence is formulated as \citep{Bertoldi92-virial,Pillai2011}
\begin{equation}\label{eq:sig_Alfven}
\sigma_{\mathrm{A}}=B(\mu_{0}\rho)^{- \frac{1}{2} }
\end{equation}
where $\rho$ is the density, $B$ is the magnetic field strength, and $\mu_{0}$ is the permeability of free space.
The Alfvén velocity contributes to the magnetic virial parameter as
\begin{equation}\label{eq:a_virB}
  \alpha_{\mathrm{B,vir}} = \frac{3 k_1 k_2 R}{GM_{\mathrm{cl}}}\,\left(\sigma^{2}_{\mathrm{v}}+\frac{\sigma^{2}_{\mathrm{A}}}{6}\right)
\end{equation}

For clumps with $\alpha_{\mathrm{vir}} < 2$, we can estimate the magnetic field strength needed to have $\alpha_{\mathrm{B,vir}} = 2$, as adapted from \cite{Bertoldi92-virial,Pillai2011,HuangB23-HMSC-CO}:
\begin{equation}\label{eq:B}
\begin{aligned}
B &= \sqrt{6\mu_{0}\rho\,\left(\frac{\alpha_{\mathrm{B,vir}}GM_{\mathrm{cl}}}{3k_1 k_2 R}-\sigma^{2}_{\mathrm{v}}\right)}.
\end{aligned}
\end{equation}

\section{Results and Discussion} \label{sec:discus}
\subsection{Overall properties and Mach numbers}
\begin{figure*} %[hp!]
% \vspace{-1cm}
\centering
% (a)
\includegraphics[width=0.4\linewidth]{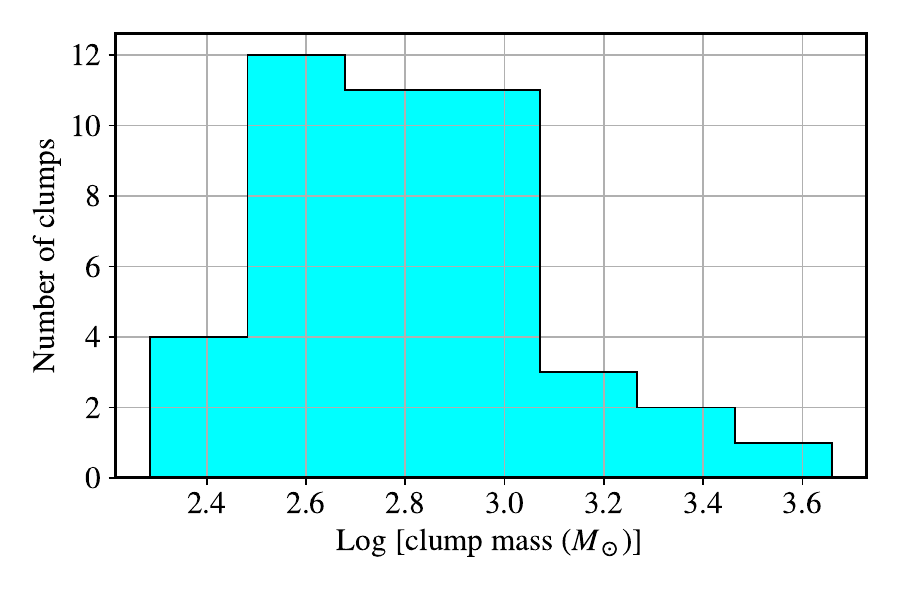}
% \vspace{-.5cm}
% (b)
\includegraphics[width=0.4\linewidth]{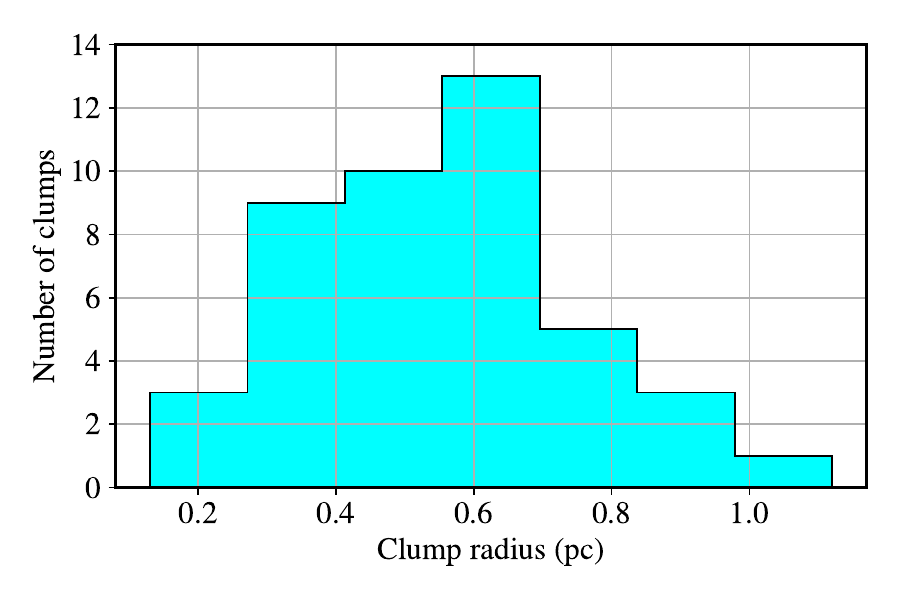}
% \vspace{-.5cm}
% (c)
\includegraphics[width=0.4\linewidth]{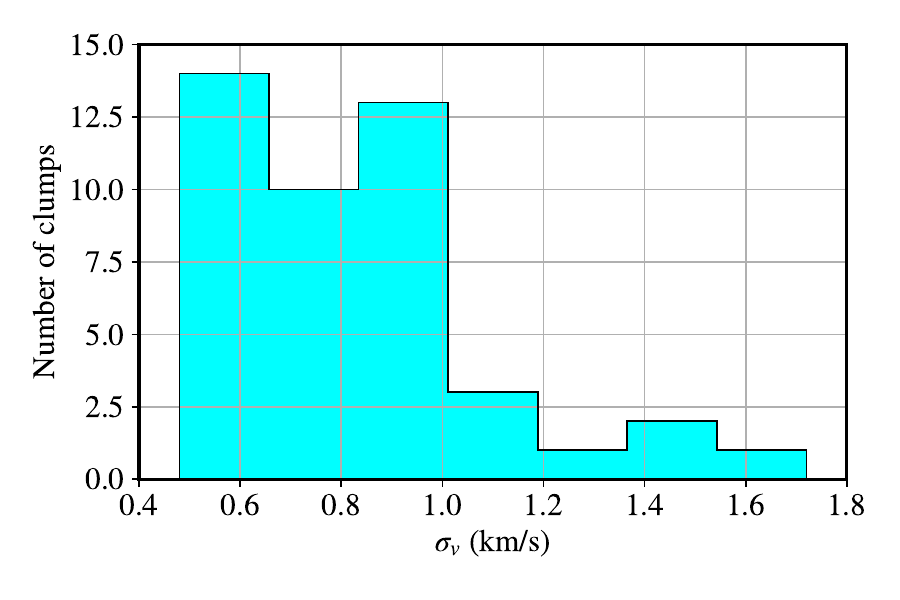}
% (d)
\includegraphics[width=0.4\linewidth]{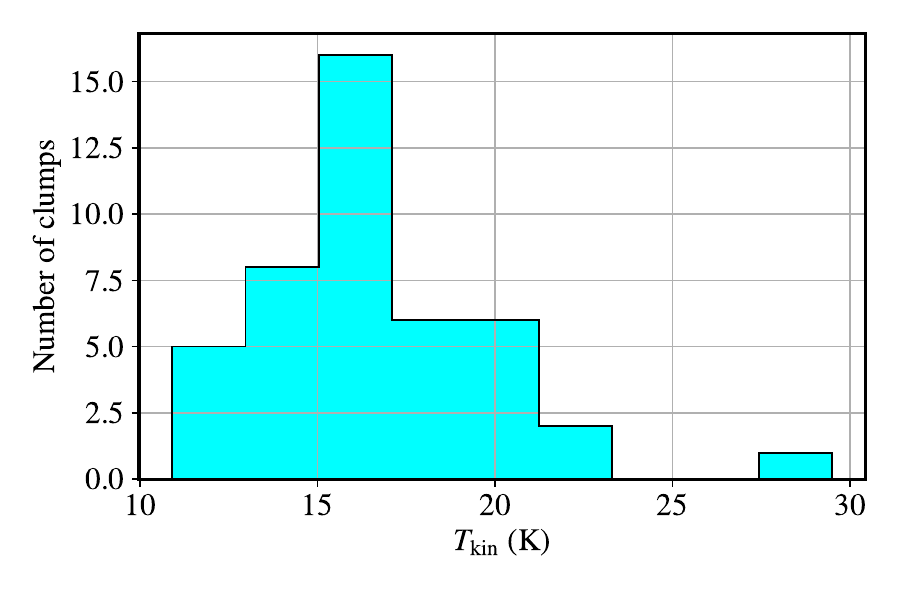}
% \vspace{-.5cm}
\includegraphics[width=0.4\linewidth]{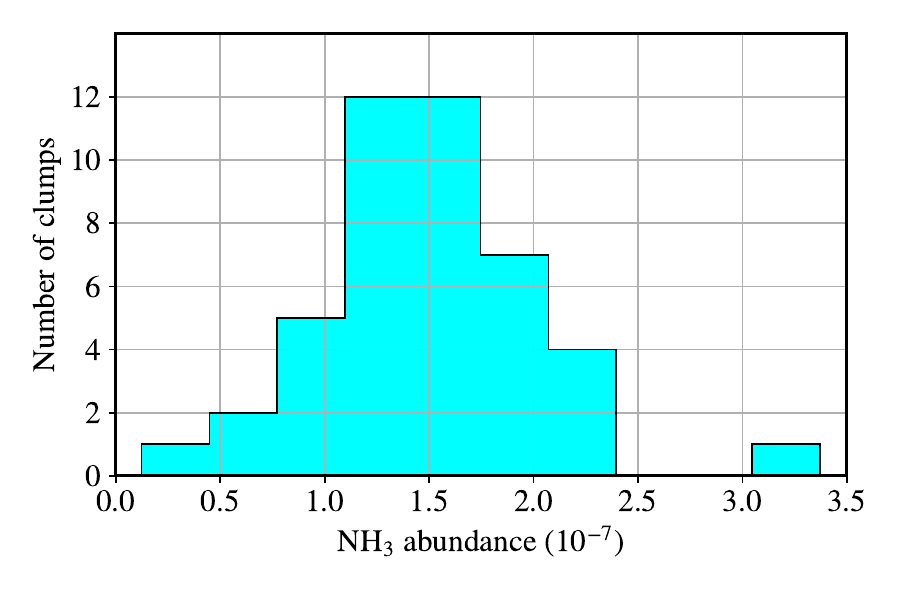}
\includegraphics[width=0.4\linewidth]{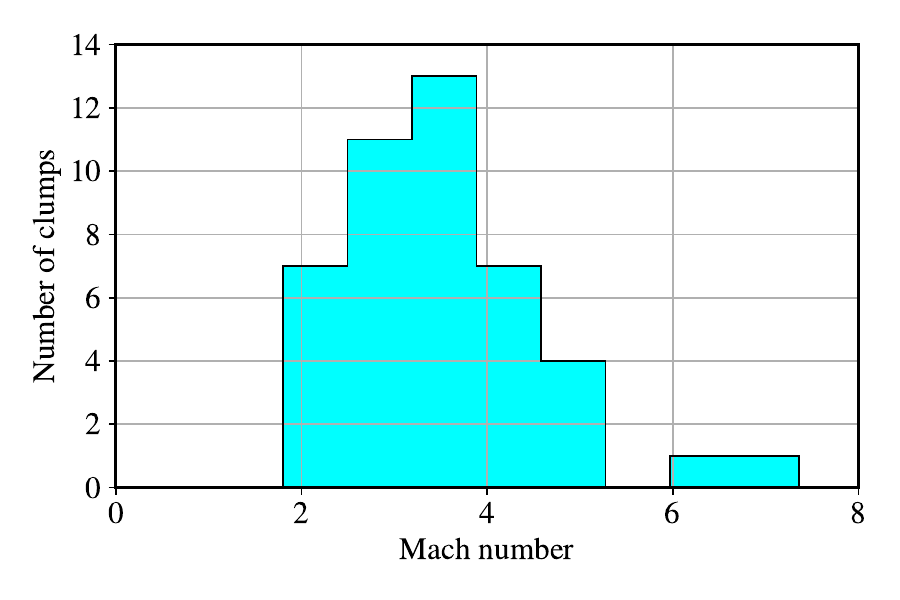}
% \vspace{-.5cm}
\includegraphics[width=0.4\linewidth]{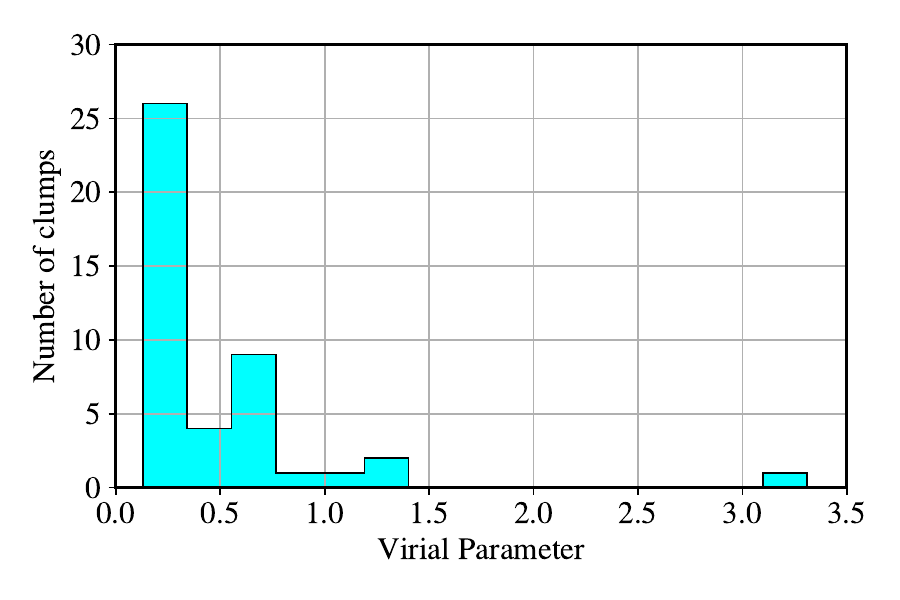}
\includegraphics[width=0.4\linewidth]{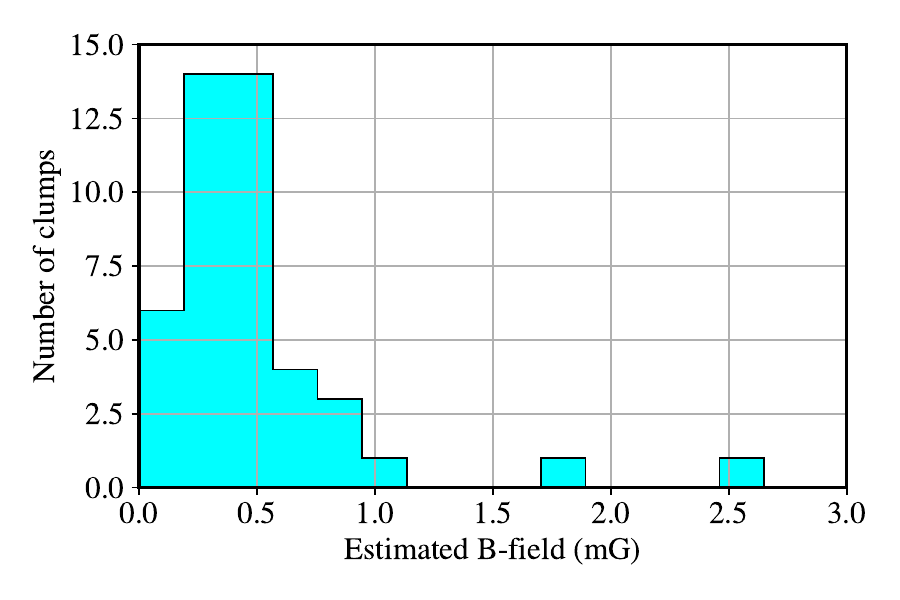}
% \vspace{-.5cm}
% \includegraphics[width=0.4\linewidth]{figure/plt_TkTd.pdf}
% \vspace{-.3cm}
% \vskip -.5cm
\caption{
Histograms of basic, \nh3\ fitted, and derived parameters of the HMSCs (Table 1): 
% Histograms show 
clump mass, equivalent radius; 
\nh3\ fitted velocity dispersion, kinetic temperature, and
\nh3\ abundance;
derived Mach number, virial parameter, and estimated magnetic field strength (for the 43 HMSCs with $\alpha_{\rm vir}<2$).
% Clump masses and radii are adopted from \cite{YuanJH17-HMSCs}, with few update, see Table 1 caption.
% The last panel shows a comparison between kinetic and dust temperature.
}
\label{fig:hist}
\end{figure*}

Table 1 lists physical properties of the HMSCs, including basic, fitted, and derived parameters, along with simple statistics.
\autoref{fig:hist} illustrates the parameters.

Median values of selected parameters are:
clump mass 593\msun,
equivalent radius 0.56\,pc,
molecular hydrogen number density $1.42\times 10^4$\cmc,
kinetic temperature 16.3 K,
velocity dispersion 0.76\kms,
Mach number 3.32,
virial parameter 0.32,
estimated magnetic field strength 0.39 mG,
and \nh3\ abundance $1.47 \times 10^{-7}$. The median values represent typical properties of the 44 HMSCs.

The Mach numbers show supersonic non-thermal motions at clump scales, but most HMSCs have \machs$\,<5$, inconsistent to what have been considered in {some theoretical works} (\autoref{sec:intro}).

\subsection{Correlation matrix}
\begin{figure*} %[hp!]
% % Answer: [trim={left bottom right top},clip]
% \includegraphics[width=.4\textwidth, trim={0 0 1cm 1.5cm},clip]{fig/LV.pdf}
\centering
% \includegraphics[width=\linewidth]{figure/corr_matrix.pdf}\\
% (a)\\
% \includegraphics[width=.49\linewidth]{figure/plt_Bn4.pdf}
% \includegraphics[width=.49\linewidth]{figure/plt_TkTd2.pdf}\\
% (b) \hspace{7.5cm} (c)
\includegraphics[width=.76\linewidth,trim={4.55cm .8cm 2.5cm .5cm},clip]{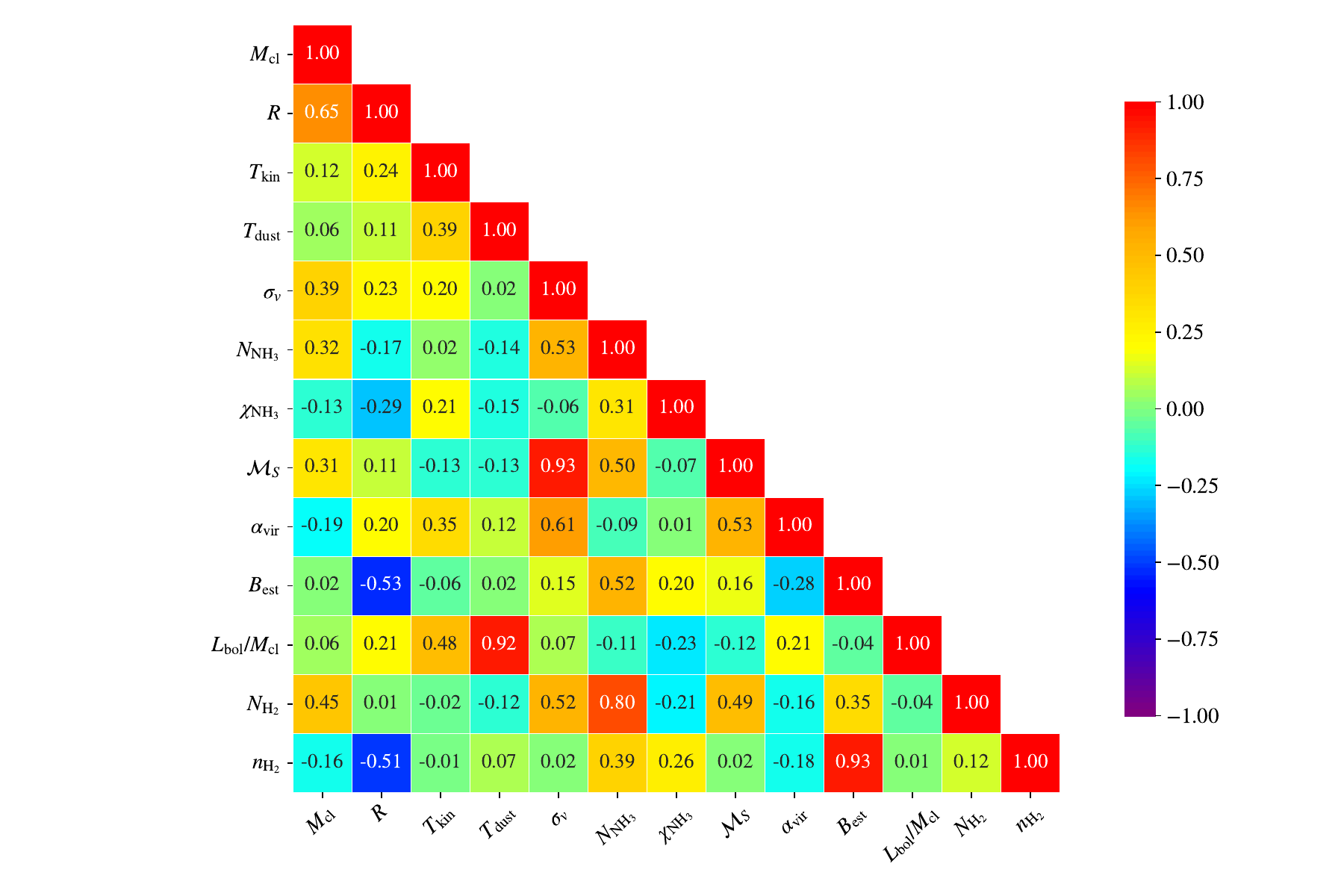}\\
(a)\\
\includegraphics[width=.66\linewidth,trim={.7cm .6cm .56cm .3cm},clip]{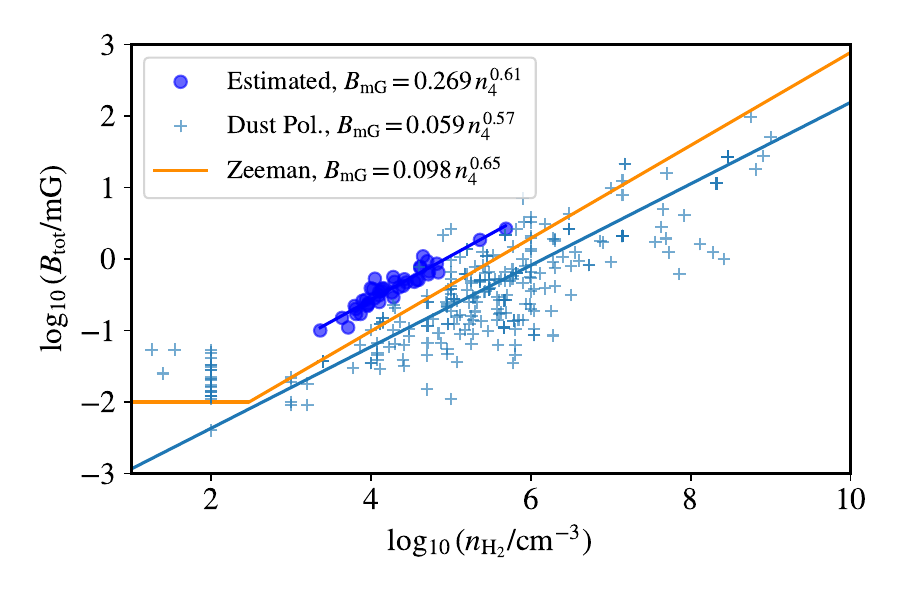}
\includegraphics[width=.33\linewidth,trim={2.5cm 0 3cm .5cm},clip]{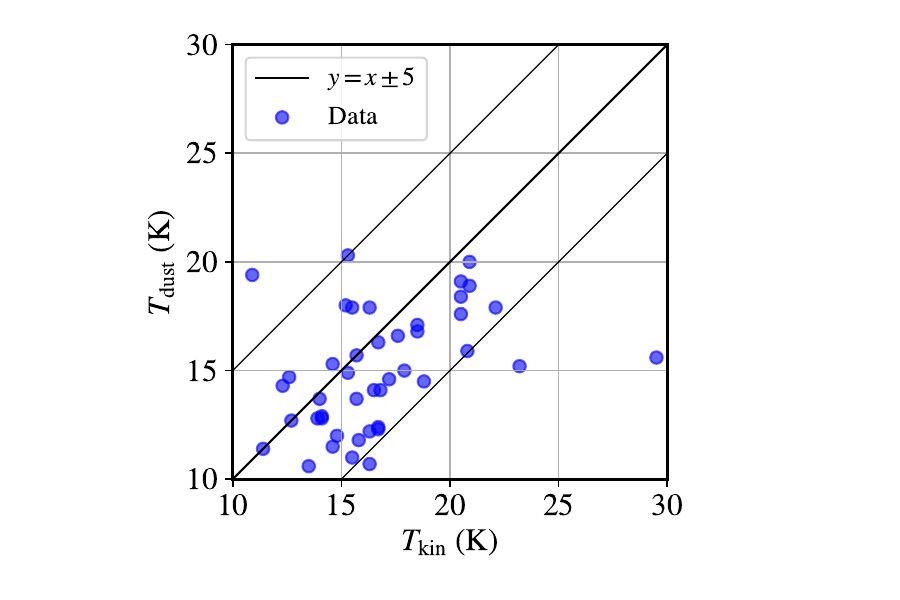}\\
\hspace{4cm} (b) \hspace{8cm} (c)
\caption{
(a)
Pearson correlation coefficient ($P$) matrix for the HMSCs. $P$ is in range [-1, 1], where $|P|>0.6$ is considered to show a strong correlation.
(b)
Comparing the estimated B-field strength to observations.
The $B_{\rm est}-n$ relation (blue circles and line) for the HMSCs are fitted as
$B_{\rm mG}=0.269\,n_4^{0.61}$, 
where $B_{\rm mG}$ is $B$ in units of mG and $n_4$ is $n_{\rm H_2}$ in $10^4$\cmc.
The 288 dust polarization measurements complied by \cite{LiuJH2022_Pol288} are shown in crosses with a fitted line in light blue. The orange line is a fit to Zeeman splitting measurements, with a break at $n_{\rm H_2} = 300$\cmc, adopted from \cite{Crutcher2010_Zeeman}.
(c)
The discrepancy between gas kinetic temperature and dust temperature. \Tdust\ is from \cite{YuanJH17-HMSCs}, which is derived from SED fitting to \her/HiGAL 70-500\um\ continuum images
% (after background/foreground subtraction) 
at a resolution of 36.4$''$, comparable to the GBT \nh3\ data.
}
\label{fig: matrix}
\end{figure*}

We explore possible correlations and trends in the parameters by building a matrix of correlation coefficients in \autoref{fig: matrix}a.\footnote{To efficiently read the matrix (\autoref{fig: matrix}a), for example, for \Best, first start with \Best\ on the left column and move to rightmost, then turn down to move vertically.}
Strong correlations are observed in three pairs of parameters: 
\begin{enumerate}
    \item estimated magnetic field strength \Best\ and volume density \nmH\ (correlation coefficient $P=0.93$);
    \item velocity dispersion \sigv\ and Mach number \machs\ ($P=0.93$);
    \item dust temperature \Tdust\ and luminosity-to-mass ratio $L/M$ ($P=0.92$).
\end{enumerate}
% Moderate correlations are shown in another three pairs of parameters:
% \begin{enumerate}
% \setcounter{enumi}{3}
%     \item \nh3\ column density and total column density ($P=0.80$);
%     \item clump radius and mass ($P=0.65$);
%     \item velocity dispersion \sigv\ and virial parameter \avir\ ($P=0.61$).
% \end{enumerate}

The $B-n$ correlation is expected theoretically \citep{Crutcher2010_Zeeman}, and the \Best\---\nmH\ correlation has been reported in \cite{HuangB23-HMSC-CO} with a weaker $P$. \Best\ also tends to be stronger in smaller clumps ($P=-0.53$), motivating interferometric observations for high resolution dust polarization \citep{LiuJH2022_Pol288,Beuther2024_Pol20,qz14_SMApol}. The \Best\---\nmH\ correlation can be fitted in \autoref{eq:Bn} (\autoref{fig: matrix}b). See more discussion in \autoref{sec:sub-viral}.
The \Tdust\---$L/M$ correlation has been shown in \cite{Elia2016_analyticSED} analytically, and has also been observed by \cite{HuangB23-HMSC-CO} in a sample of 207 HMSCs.
The \sigv\---\machs\ correlation is evident in \autoref{eq:mach}, with little span in sound speed in these HMSCs.

\autoref{fig: matrix}a
also reveals no clear correlation in a few notable pairs of parameters.
Firstly, the dust and gas temperature are not well correlated, although the difference among the two are mostly $<$5\,K (\autoref{fig: matrix}c). 
% An outlier ...
Gas temperature has a larger dynamical range (10.9--29.5 K) than dust temperature (10.6--20.3 K), as have been pointed out in \cite{me2018_G28}. %Merello2019
Poor correlation between dust and gas temperature further necessaries temperature derived using spectral lines with radial velocity corresponding to the source's emission, to avoid line-of-sight contamination. \cite{YuanJH17-HMSCs} have already subtracted a modeled background/foreground before fitting SED. Still, our results show that spectral line derived gas temperature is necessary to better constrain the temperature. {See more discussion on \Tdust\ and \Tkin\ in \cite{me2018_G28,Merello2019}.}
% {Could Tdust Tgas intrsinsically different? Moller2019.}
Secondly,
clump radii do not show clear correlation with Mach number or velocity dispersion, although positive trends ($P>0$) are evident. It is known that turbulence dissipates from large to small scales in a given molecular cloud \citep[e.g.][]{WangY2008,YueNN2021_linewidth}. The observational results here indicate that turbulence can be different from one source to another, and multi-scale observations in individual sources are needed to further investigate the dissipation of turbulence over spatial scales \citep{LiuJH2024_G28}. 
Thirdly, \nh3\ fractional abundance does not appear to correlate with evolutionary indicators like $L/M$ and temperature, suggesting that \nh3\ abundance is stable in the initial stages as in HMSCs.
Lastly, the linewidth-size relation \citep{Larson81-turbulent} does not hold in scales smaller than $\lesssim$1 pc, as already found in a number of observations \citep[e.g.][]{me2009,HuangB23-HMSC-CO}.
{Notably, some of the non-correlations may be partly affected by narrow ranges of parameters, as the HMSCs are strictly selected to be starless. This needs to be taken into account particularly when comparing these (non-)correlations with those derived from clumps in a wider evolutionary stages, and therefore occupying larger parameter ranges. }

\subsection{Sub-{virial} clumps and implications} \label{sec:sub-viral}
The most intriguing finding is the small \avir, which is in range 0.13-1.37 except an outlier at 3.31 (G030.7912-0.1173). The outlier source has a marginal detection in \nh3\ (2,2), and the fitted line velocity dispersion is $\sigma = 1.72\pm0.48$\kms, with a large uncertainty compared to other sources.
% outflow? Yang22
So, all but one of the 44 HMSCs have \avir\,$<2$ (40 of them have \avir\,$<1$), suggesting that internal kinetic energy is insufficient to support HMSCs against from gravitational collapse.
Similarly small virial parameters have been reported by \cite{Merello2019} in a comparative study by cross-matching HiGAL clumps to literature \nh3\ studies. Unlike the uniform RAMPS data, the literature data have diverse sky coverages and various spectral resolution, some of which are larger than sound speed. In a classic Effelsberg \nh3\ study of IRDCs, \cite{Pillai2006} find \avir$\sim$1.

{\cite{Kauffmann13-Avir} compiled and standardized from the literature 1325 virial parameter estimates on a range of cloud structures, from clouds, clumps, to cores. They reported low virial parameters of $\ll$2 in some high-mass star formation regions, mostly from the ATLASGAL clumps of \cite{Wienen2012}. A close inspection of their Fig.\,1 finds that our virial parameters (median 0.32, mostly less than 0.7) populate the lower end of the \cite{Wienen2012} virial parameters. This is reasonable, because most of the \cite{Wienen2012} clumps are of protostellar nature (see Introduction), and star formation activity within the clumps have broadened the initial linewidth.}

In order to stabilize the 43 HMSCs with \avir\,$<2$, magnetic fields with strength of 0.10-2.65 mG (median 0.39 mG) would be needed (Table 1).
We compare the estimated strengths to that derived from observations at the clump scales. 
% \cite{Pillai15-Bfield} analyzed two infrared dark clouds (IRDCs), G11.11-0.12 and G0.253+0.016 using dust polarization at 850 and 350\um, and found 0.267\,mG and 5.432\,mG, respectively.
% 850 $\mu \mathrm{m}$ polarization data and 350 $\mu \mathrm{m}$ archival polarization data, respectively. They calculated the total magnetic field $B_{\mathrm{tot}}$, finding values of 0.267 mG for the former and 5.432 mG for the latter. 
% \cite{Liu22-Bfield} analyzed a sample of sources using the Davis-Chandrasekhar-Fermi (DCF) method to estimate the strength of the B fields. From the 288 measurements, the total strength of the B field ranges from 0.01 to 10 mG.
\cite{LiuT2018_G35.39pol} present JCMT 850\um\ dust polarization observations towards a well characterized filamentary IRDC G035.39-00.33 \citep[e.g.,][]{Izaskun2010_G35.39_SiO,Henshaw16-velo-disper,Sokolov2017,Sokolov18-velo-disper,Sokolov2019}.
At a resolution of 14$''$ (0.18\,pc), the observations measure a mean plane-of-the-sky magnetic field strength of $\sim$0.05\,mG, much lower than \Best\ in the HMSCs even assuming a similar line-of-sight strength. The 6.8\,pc main filament as a whole is gravitationally unstable if it is only supported by thermal pressure and turbulence; adding the magnetic support can stabilize the northern part of the filament, but not the southern and central parts. Nine clumps are distributed along the filament, with mass 35-219\msun, effective radii 0.12-0.31\,pc, and $B_{\rm clump}$ 0.056-0.219\,mG. The clumps are similar to the smallest HMSCs in Table 1, and the observed B-field strengths are among the smallest compared to \Best. Considering that \Best\ tends to become stronger in smaller HMSCs ($P=-0.53$, \autoref{fig: matrix}a), the required \Best\ to stabilize HMSCs would be strong compared to the observations in G035.39-00.33.

\cite{LiuJH2022_Pol288} compiled 288 polarized dust emission observations from the literature and recalculated the total magnetic field strength using a revised Davis-Chandrasekhar-Fermi (DCF) method. In a classic $B-n$ plot (\autoref{fig: matrix}b), the 288 measurements are compared with the estimated B-field strength in the 43 HMSCs. The \Best\ are systematically higher than observed strength in the same density range.
The \Best\ shows a tight correlation with density, which can be fitted as
\begin{equation} \label{eq:Bn}
B_{\rm mG}=0.269\,n_4^{0.61}
\end{equation}
where $B_{\rm mG}$ is $B$ in units of mG and $n_4$ is $n_{\rm H_2}$ in $10^4$\cmc.
\cite{LiuJH2022_Pol288} fitted $B_{\rm mG}=0.059\,n_4^{0.57}$ for polarized dust measurements, and \cite{Crutcher2010_Zeeman} fitted $B_{\rm mG}=0.098\,n_4^{0.65}$ for Zeeman splitting measurements (at $n_{\rm H_2}>300$\cmc).
The power-law indices are similar but the strengths are different.
For the typical density of $1.42\times 10^4$\cmc\ of the HMSCs, \Best\ is 4.6 and 2.7 times that of the polarized dust measurements and Zeeman splitting measurements, respectively.
It is important to note that the $B$ observations are mainly from star formation regions at later evolutionary stages than HMSCs.  Observations towards the initial stages \citep[e.g.][]{LiuT2018_G35.39pol} are very rare so far. 

% We compile all the previous DCF estimations from polarized dust emission observations and recalculate the magnetic field strength of the selected samples with the new DCF correction factors in 

% \cite{Stephens2022_G47_SOFIA}
% SOFIA G47:
% strument at 214 μm and 18.″2 resolution.
% vary from ~20 to ~100 μG.

% \cite{Soam2019_G34.43Pol}
% G34.43+0.24
% We obtained a plane-of-sky magnetic field strength of 470 ± 190 μG, 100 ± 40 μG, and 60 ± 34 μG in the central, northern, and southern regions of G34, respectively, using the updated Davis-Chandrasekhar-Fermi relation. 

The results suggest that global dynamic collapse at clump scale is common, which is widely observed in massive clumps with global infall signatures \citep{HeYX2015_infall,HeYX2016_infall,YuSL2022_infall_cat,XuFW2023_JCMT_infall_mapping}
and resolved multiscale gas flows \citep[e.g.,][]{YuanJH2018_G22, ZhouJW2022_HFS_infall,ZhangSJ2023_ClumpFed_Triggered,Xu2023SDC335}.

The implications of these findings are two folds.
Observationally, constraints on the \textit{initial} conditions of massive star formation have finally reached a Galaxy-wide sense, to a pilot yet representative sample of 44 HMSCs built on our previous work \citep{YuanJH17-HMSCs,HuangB23-HMSC-CO}. These HMSCs represent the very first observed physical properties that are important to refine the initial inputs to theoretical models. 
% The systematic characterization of \cite{HuangB23-HMSC-CO} suffered from overestimation of kinetic energy using $^{13}$CO\,(3--2) linewidth.
Theoretically, the prevalence of sub-viral clumps {(radii 0.13-1.12 pc, Table 1)} is in stark contrast to the virial equilibrium assumed in the turbulent core model (\autoref{sec:intro}). 
{Note that in this model both cores and clumps are assumed to be in virial equilibrium, based on limited observational constraints available at that time \citep[see more details in][]{Mckee03-formmodel}.
This assumption only applies to the initial conditions, i.e., starless/prestellar stage, not protostellar stage \citep{Mckee03-formmodel}. Sub-virial presteallar and protostellar cores (smaller than most of the HMSCs) have been reported \citep[e.g.,][]{ZhangQZ15-HMSCs,Sanhueza2017,LiuJH2020_G28pol,Barnes2021,LiSH2023,JiaoWY23-HMSCs}.}
The findings in this work are drawn from a systematic characterization using uniform data on a uniform sample of HMSCs, providing the first statistically significant constraints on the \textit{{initial}} conditions. The HMSCs are therefore of great interest for follow-up observations at high-resolution, to resolve the beginning of star formation \citep[e.g.,][for G010.2144-0.3051]{JiaoWY23-HMSCs}. 
% For example, \cite{JiaoWY23-HMSCs} present ALMA and SMA observations of one of the HMSCs, G010.2144-0.3051.

% Moreover, 
The findings urge future observations to quantify the support of magnetic fields in the initial, starless stages of massive star formation regions, like the HMSCs well characterized here. Although challenging to observe, dust polarization imaging by JCMT, SOFIA, SMA, and ALMA {has recently been carried out toward} massive clumps bright at sub-millimeter, mostly at a later evolutionary stage than HMSCs \citep[e.g.][]{qz14_SMApol,Beuther2024_Pol20,Sanhueza2021_I18089_ALMApol,Stephens22-SOFIA_G47Pol,LiuJH2022_Pol288,Pattle2023-PP7}. 
% \citep{Stephens2022_G47_SOFIA,LiuJH2020_G28pol,LiuJH2022_Pol288,LiuT2018_G35.39pol,Soam2019_G34.43Pol,LiuJH2024_G28,Beuther2024_Pol20,Beltran2024_G34.41pol,qz14_SMApol} QZ and Chat Hull's review
It would be even more challenging to do so for HMSCs because they are generally less bright. {If the B-field strengths in HMSCs are indeed as strong as \Best\ in \autoref{fig: matrix}(b), observations may be less time consuming than previously thought.}
% {The strong estimated B-field strength in Fig 4b thus give a convenient constraint:
% if xxx, then much easier to observe;
% or, not so strong, then theory down.
% G11 IR pol.}
Alternatively, velocity fields and density gradients have been proposed to indirectly probe magnetic fields \citep[e.g.,][]{Jiao2024_OrionA}.
%(Hu, Jiao, Planck) 
The only direct measurement of the magnetic field strength relies on observing Zeeman splitting {\citep{Crutcher2010_Zeeman,Crutcher2012}}. Recently, \cite{Ching2022_Nature} have demonstrated the feasibility of using HI narrow self-absorption (HINSA) to measure Zeeman splitting with FAST. {Several well-characterized HMSCs here} can be targets for such exciting experiments in the FAST era, which we are carrying out \citep{SunSL2024}.

{The huge difference between the estimated B-field strengths needed to stabilize starless clumps and that observed in protostellar and more evolved clumps strongly suggests a dynamical star formation initiated by collapsing, not equilibrium, clumps. Without sufficient support from magnetic fields, the collapse would be very quick, nearly free-fall, resulting in short lifetime of the high-mass starless cores in the framework of the turbulent core model \citep{Kauffmann13-Avir}. This is consistent with the very few numbers of observed high-mass starless cores after decades of searching \citep[e.g.,][]{Bontemps2010,WangK14-HMSCs,Motte18-Starform,JiaoWY23-HMSCs,JiaoWY2024_OrionStarless,Barnes2023}.}

\section{Conclusions}\label{sec:conc}

% The initial conditions 
% are critical for understanding high-mass star formation, but are not well observed. 
Built on our previous Galaxy-wide characterization of $\sim$pc scale high-mass starless clumps (HMSCs), here we have investigated the dynamical state of a representative subsample of 44 HMSCs, aiming to reveal the initial conditions for high-mass star formation. {The main} findings and implications are as follows.

% \begin{enumerate}

% \item 
(1)
By modeling GBT \nh3\ (1,1) and (2,2) spectra extracted from each of the HMSCs, we derived 
line velocity dispersion (range 0.48-1.72, median 0.76\kms), 
gas kinetic temperature (range 10.9-29.5, median 16.3 K), 
\nh3\ column density ($1.23\times10^{-8} - 3.37\times10^{-7}$, median $1.47\times10^{-7}$), 
Mach number (1.80-7.36, median 3.32, mostly $<$5), 
and virial parameter (range 0.13-1.37, median 0.32, and an outlier at 3.31).

% \item 
(2)
The Mach numbers show that these HMSCs have supersonic turbulence, but most of them have \machs\ lower than 5, inconsistent to what have been considered in {some theoretical works} (\autoref{sec:intro}). 

% \item 
(3)
Thermodynamic analysis reveals that all but one (43/44) of the HMSCs are gravitationally bound with virial parameter $\alpha_{\mathrm{vir}} < 2$.
In order to balance these HMSCs from gravity, as suggested in the turbulent core model, the 
% Either these massive clumps are in collapsing or magnetic field strengths of 0.10-2.65\,mG (average 0.51\,mG) would be needed to support them against collapse. 
% The 
estimated B-field strength correlates tightly with density,
$B_{\rm est}/{\rm mG}=0.269\,(n_{\rm H_2}/10^4\,{\rm cm^{-3}})^{0.61}$,
with a similar power-law index as found in previous dust polarization observations, but a factor of 4.6 higher in strength.

% \item 
(4)
For the first time, the initial dynamical state of high-mass formation regions has been constrained to be sub-virial in a statistically significance manner with uniform data and sample. This is in contradictory to theoretical models in virial equilibrium, {while} in agreement with recent observations revealing a dynamic, not quasi-static, massive protocluster formation, {and consistent with the lack of observed high-mass starless cores.}

% \item 
(5)
The findings urge future observations to quantify the magnetic field support in the prestellar stage of massive clumps, which are rarely explored so far, towards a full understanding of the physical conditions that initiate massive star formation.
The Five-hundred-meter Aperture Spherical radio Telescope (FAST) can make a significant contribution with demonstrated Zeeman splitting observations of HINSA, to directly measure magnetic field strength.

The HMSCs well characterized here represent the very first observed properties that would initiate massive star formation, if not already. These properties should be accounted in theoretical explorations. Further ALMA observations are ongoing to resolve the beginning of massive star formation within these HMSCs \citep[e.g.,][]{JiaoWY23-HMSCs}.  

% \end{enumerate}

\begin{acknowledgements}
{We thank the referee for in-depth evaluation and constructive suggestions that helped strengthen the original manuscript.}
We acknowledge support from the National Natural Science Foundation of China (12041305, 12033005), 
% the China Manned Space Project (CMS-CSST-2021-A09, CMS- CSST-2021-B06), the National Key Research and Development Program of China (2017YFA0402702, 2019YFA0405100), 
the Tianchi Talent Program of Xinjiang Uygur Autonomous Region, the China-Chile Joint Research Fund (CCJRF No. 2211),
and the High-Performance Computing Platform of Peking University.
CCJRF is provided by Chinese Academy of Sciences South America Center for Astronomy (CASSACA) and established by National Astronomical Observatories, Chinese Academy of Sciences (NAOC) and Chilean Astronomy Society (SOCHIAS) to support China-Chile collaborations in astronomy.
This publication makes use of molecular line data from the Radio Ammonia Mid-Plane Survey (RAMPS). RAMPS is supported by the National Science Foundation under grant AST-1616635.
{The Green Bank Observatory is a facility of the National Science Foundation operated under cooperative agreement by Associated Universities, Inc.}
\end{acknowledgements}

\setlength\tabcolsep{3.5pt}

\begin{longrotatetable}
\begin{deluxetable*}{lcccc ccccc ccccc cccc}
% {lrrrr rrrrr rrrrr rrrrr rrrrrr}
% Name                 ra       dec     velo  Dist  Tkin  sigV  FWHM    N_nh3   Mach  avir  Best  Tdust   Mcl      LM   r_pc    Nh2      ab_NH3    n_H2  
\tabletypesize{\scriptsize}
% \tabletypesize{\tiny}
\tablewidth{5cm} 
%\tablenum{1}
\tablecaption{Physical Properties of the 44 HMSCs \label{tab:hmsc44}}
\tablehead{
\colhead{Name} &
\colhead{RA} &
\colhead{Dec} &
\colhead{\vlsr} &
\colhead{$d$} &
\colhead{\Tkin} &
\colhead{\sigv} &
\colhead{$\Delta V_{\rm FWHM}$} &
\colhead{$N_{\rm NH_3}$} &
\colhead{\machs} &
\colhead{\avir} &
\colhead{\Best} &
\colhead{\Tdust} &
\colhead{$M_{\rm cl}$} &
\colhead{$L/M$} &
\colhead{$R$} &
\colhead{$N_{\rm H_2}$} &
\colhead{$\chi_{\rm NH_3}$} &
\colhead{$n_{\rm H_2}$} \\
\colhead{} &
\colhead{deg} &
\colhead{deg} &
\colhead{\kms} &
\colhead{kpc} &
\colhead{K} &
\colhead{\kms} &
\colhead{\kms} &
\colhead{ } &
\colhead{ } &
\colhead{ } &
\colhead{mG} &
\colhead{K} &
\colhead{\msun} &
\colhead{\lsun/\msun} &
\colhead{pc} &
\colhead{\cms} &
\colhead{ } &
\colhead{\cmc} 
} 
\colnumbers
\startdata 
\input{tab_data}
\enddata
\tablecomments{ %\small
% Name                 ra       dec     velo  Dist  Tkin  sigV  FWHM    N_nh3   Mach  avir  Best  Tdust   Mcl      LM   r_pc    Nh2      ab_NH3    n_H2  
Col. (1) HMSC name.
Col. (2-3) J2000 coordinates.
Col. (4) \nh3\ fitted radial velocity.
Col. (5) Distance.
Col. (6) kinetic temperature.
Col. (7-8) velocity dispersion and FWHM linewidth.
Col. (9) \nh3\ column density.
Col. (10-12) Derived parameters: Mach number, virial parameter, and estimted magnetic field strength.
Col. (13-17) Dust temperature, clump mass, luminosity-to-mass ratio, equvalent radius, column density.
Col. (18) \nh3\ abundance.
Col. (19) volume density.\\
Data for Col. (1-3, 5, 13-17, 19) are adopted or updated from \cite{YuanJH17-HMSCs}.
The \nh3\ fitted \vlsr\ is consistent with that of \cite{YuanJH17-HMSCs}, except in one source, G030.7941+0.0736. It's \vlsr\ has been updated to 90.0\kms\ according to \nh3\ detection. \cite{YuanJH17-HMSCs} adopted 36.4\kms\ from literature. Both of the two \vlsr\ correspond to velocity components along the line of sight, and 90.0\kms\ coincides with the brightest $^{13}$CO\,(3--2) peak in the JCMT spectra \citep{Dempsey2013_JCMT_CO32}. The kinematic distance is updated using \vlsr$=90$\kms, and related parameters are updated accordingly.
% https://atlasgal.mpifr-bonn.mpg.de/cgi-bin/ATLASGAL_SEARCH_RESULTS.cgi?text_field_1=AGAL030.794%2B00.074&catalogue_field=Sextractor&gc_flag=
The derived \Tkin\ is not always consistent with \Tdust. Assuming coupling between dust and gas, and because dust temperature may suffer from line-of-sight contamination, we have updated the mass by using \Tkin\ instead of the \Tdust, in a Rayleigh-Jeans approximation, i.e., mass scales linearly with temperature.
}
\end{deluxetable*}
\end{longrotatetable}

% \begin{figure*}[ht!]
% \centering
% \includegraphics[width=0.9\linewidth]{figure/overview_yuan.pdf}
% \caption{Overview of an example high-mass starless clump (HMSC) G022.5309-0.1927.
% % environments with multiwavelength maps and the RAMPS \nh3\ moment-0 map. 
% (a) Three pseudo-color images with \textit{Spitzer} emission at 8.0, 4.5, and 3.6 $\mu$m rendered in red, green, and blue, respectively. (b) \textit{Spitzer} MIPS 24 $\mu$m emission. (c) \textit{Herschel} 70 $\mu$m emission. (d) \textit{Herschel} 250 $\mu$m emission. (e) the dust temperature shown in background and H$_2$ column density in gray contours of [2.0, 2.5, 3.0, 3.5, 4.0, 4.5, 5.0]$ \times 10^{22}$ cm$^{-2}$. (f) moment 0 map of $\mathrm{NH_{3}}$(1,1) data of RAMPS show a good spatial correlation with cold dust emission; the GBT beam size is shown at the bottom left corner. In all panels, the white dashed ellipse outlines the source size.}
% \label{fig:rgb}
% \end{figure*}

\facilities{GBT, \textit{Spitzer}, \textit{Herschel}}
\software{{Pysepckit} \citep{Ginsburg22-Pysepckit} }

\bibliography{my.bib}{}
\bibliographystyle{aasjournal}
\end{document}

%% file: mycmd.tex
\def\nh2d{$\rm{NH_2D}$}

\def\nh3{NH\textsubscript{3}}
\def\NH3{$\rm{NH_3}$}

\def\n2hp{N\textsubscript{2}H\textsuperscript{+}}

\def\h2o{$\rm{H_2O}$}
\def\h2{$\rm{H_2}$}

\def\msun{\,$M_\odot$}

\def\lsun{\,$L_\odot$}

\def\um{\,$\mu\mathrm{m}$}

\def\kms{\,km~s$^{-1}$}

\def\cm2{\,$\rm{cm^{-2}}$}
\def\cm3{\,$\rm{cm^{-3}}$}
\def\cms{\,$\rm{cm^{-2}}$}
\def\cmc{\,$\rm{cm^{-3}}$}

\def\vlsr{$V\rm{_{LSR}}$}

\def\11{(1,1)}
\def\22{(2,2)}
\def\33{(3,3)}
\def\44{(4,4)}
\def\55{(5,5)}
\def\t21{$T_{21}$}
\def\r31{$R_{31}$}

\newcommand{\Trot}{$T_{\rm rot}$}
\newcommand{\Tkin}{$T_{\rm kin}$}

\newcommand{\Tdust}{$T_{\rm dust}$}

\newcommand{\mach}{$\mathcal{M}$}
\newcommand{\machs}{$\mathcal{M}_S$}
\newcommand{\avir}{$\alpha_{\rm vir}$}
\newcommand{\sigv}{$\sigma_{\rm v}$}
\newcommand{\Best}{$B_{\rm est}$}
\newcommand{\nmH}{$n_{\rm H_2}$} % n molecular hydrogen

\newcommand{\cob}{\textsuperscript{13}CO\,(2{\textendash}1)}

\def\her{\emph{Herschel}}

\let \amp = \&

%% file: tab_data.tex
% Name                 ra       dec     velo  Dist  Tkin  sigV  FWHM    N_nh3   Mach  avir  Best  Tdust   Mcl      LM   r_pc    Nh2      ab_NH3    n_H2    
G009.9517-0.3649 & 272.2560  &-20.5094 & 13.0  & 3.12  &14.8  &0.51  &1.21  &5.67E+15  &2.21  &0.29  &0.29  &12.0  &2.65E+02  &0.33  &0.36  &4.82E+22  &1.18E-07  &1.91E+04 \\
G010.1904-0.3884 & 272.4012  &-20.3118 & 10.9  & 3.12  &20.9  &0.93  &2.18  &1.21E+16  &3.38  &0.61  &0.61  &18.9  &3.15E+02  &4.48  &0.28  &6.34E+22  &1.91E-07  &5.21E+04 \\
G010.1976-0.2876 & 272.3108  &-20.2568 & 10.5  & 3.12  &15.2  &0.51  &1.21  &4.74E+15  &2.18  &0.22  &0.65  &18.0  &2.15E+02  &2.67  &0.22  &4.01E+22  &1.18E-07  &6.97E+04 \\
G010.2144-0.3051 & 272.3358  &-20.2506 & 12.0  & 3.12  &17.6  &0.88  &2.06  &1.43E+16  &3.49  &0.27  &2.65  &16.6  &2.96E+02  &2.26  &0.13  &6.21E+22  &2.30E-07  &4.89E+05 \\
G010.2190-0.3632 & 272.3924  &-20.2746 & 12.0  & 3.12  &20.5  &0.99  &2.32  &8.46E+15  &3.64  &0.73  &0.51  &18.4  &3.33E+02  &3.67  &0.31  &6.26E+22  &1.35E-07  &3.94E+04 \\
G010.2488-0.1101 & 272.1718  &-20.1259 & 14.2  & 14.56 &18.5  &1.01  &2.37  &7.39E+15  &3.92  &0.20  &0.53  &16.8  &4.57E+03  &2.25  &1.12  &3.90E+22  &1.89E-07  &1.12E+04 \\
G022.5309-0.1927 & 278.2485  &-9.3350  & 76.6  & 4.98  &16.7  &0.95  &2.23  &8.26E+15  &3.87  &0.66  &0.28  &12.4  &6.56E+02  &0.48  &0.59  &3.85E+22  &2.14E-07  &1.10E+04 \\
G022.7215-0.2733 & 278.4099  &-9.2032  & 73.4  & 4.64  &16.7  &0.50  &1.18  &3.48E+15  &2.04  &0.21  &0.24  &16.3  &5.79E+02  &2.03  &0.60  &3.69E+22  &9.43E-08  &9.29E+03 \\
G023.2674-0.3559 & 278.7389  &-8.7567  & 77.2  & 4.52  &12.3  &1.54  &3.62  &7.66E+15  &7.36  &1.08  &0.86  &14.3  &5.39E+02  &0.65  &0.30  &4.49E+22  &1.71E-07  &6.66E+04 \\
G023.2694-0.2097 & 278.6085  &-8.6876  & 78.9  & 4.74  &15.8  &0.82  &1.92  &8.25E+15  &3.43  &0.29  &0.48  &11.8  &1.08E+03  &0.31  &0.57  &7.39E+22  &1.12E-07  &1.98E+04 \\
G023.2778-0.2149 & 278.6171  &-8.6826  & 78.4  & 4.89  &14.6  &1.02  &2.39  &1.10E+16  &4.46  &0.30  &0.93  &11.5  &1.20E+03  &0.25  &0.43  &9.14E+22  &1.20E-07  &5.07E+04 \\
G023.2826-0.2007 & 278.6066  &-8.6718  & 79.8  & 4.83  &16.3  &1.13  &2.65  &1.06E+16  &4.68  &0.67  &0.35  &10.7  &1.00E+03  &0.20  &0.65  &7.61E+22  &1.39E-07  &1.27E+04 \\
G023.2957+0.0556 & 278.3827  &-8.5420  & 55.5  & 3.63  &11.4  &0.91  &2.13  &7.34E+15  &4.50  &0.60  &0.51  &11.4  &3.55E+02  &0.21  &0.32  &4.52E+22  &1.62E-07  &3.72E+04 \\
G023.2989-0.2501 & 278.6586  &-8.6801  & 82.7  & 5.44  &16.5  &0.61  &1.45  &6.15E+15  &2.51  &0.13  &1.85  &14.1  &4.52E+02  &0.90  &0.19  &3.03E+22  &2.03E-07  &2.31E+05 \\
G023.3222-0.0775 & 278.5145  &-8.5799  & 97.1  & 5.82  &14.0  &0.74  &1.73  &6.00E+15  &3.28  &0.27  &0.52  &13.7  &7.63E+02  &0.62  &0.46  &4.17E+22  &1.44E-07  &2.62E+04 \\
G023.4556-0.2301 & 278.7135  &-8.5317  & 103.6 & 5.93  &15.3  &0.48  &1.14  &3.51E+15  &2.04  &0.19  &0.20  &20.3  &7.05E+02  &5.98  &0.72  &2.72E+22  &1.29E-07  &6.45E+03 \\
G023.4742+0.1037 & 278.4227  &-8.3614  & 87.0  & 5.57  &16.8  &1.00  &2.35  &1.04E+16  &4.07  &0.34  &0.78  &14.1  &1.05E+03  &0.94  &0.44  &6.08E+22  &1.71E-07  &4.14E+04 \\
G023.4771+0.1147 & 278.4142  &-8.3538  & 87.6  & 5.56  &16.3  &0.74  &1.73  &6.76E+15  &3.04  &0.20  &0.75  &12.2  &9.16E+02  &0.40  &0.42  &5.47E+22  &1.24E-07  &4.14E+04 \\
G023.5176+0.2425 & 278.3185  &-8.2589  & 82.4  & 5.40  &15.5  &0.64  &1.52  &1.90E+15  &2.73  &0.32  &0.48  &17.9  &3.39E+02  &2.74  &0.32  &2.15E+22  &8.85E-08  &3.49E+04 \\
G023.9790+0.1498 & 278.6164  &-7.8922  & 82.3  & 4.43  &15.7  &0.66  &1.57  &5.92E+15  &2.79  &0.32  &0.42  &15.7  &4.37E+02  &1.57  &0.39  &4.10E+22  &1.44E-07  &2.52E+04 \\
G024.0179+0.2107 & 278.5800  &-7.8296  & 106.5 & 5.98  &15.3  &0.88  &2.06  &5.21E+15  &3.75  &0.51  &0.22  &14.9  &1.02E+03  &1.08  &0.82  &4.20E+22  &1.24E-07  &6.25E+03 \\
G024.0492-0.2145 & 278.9756  &-7.9976  & 81.6  & 4.80  &12.7  &0.72  &1.68  &5.08E+15  &3.35  &0.57  &0.17  &12.7  &4.53E+02  &0.47  &0.62  &3.40E+22  &1.49E-07  &6.52E+03 \\
G024.2583+0.1022 & 278.7889  &-7.6662  & 99.7  & 5.89  &16.7  &0.55  &1.30  &5.19E+15  &2.25  &0.31  &0.25  &12.3  &3.79E+02  &0.46  &0.47  &2.18E+22  &2.38E-07  &1.27E+04 \\
G024.4326+0.3238 & 278.6714  &-7.4094  & 114.9 & 6.81  &23.2  &1.18  &2.77  &2.68E+15  &4.09  &1.37  &0.15  &15.2  &6.63E+02  &1.88  &0.81  &1.99E+22  &1.35E-07  &4.35E+03 \\
G028.1943-0.0747 & 280.7620  &-4.2496  & 98.6  & 8.19  &14.6  &0.67  &1.59  &5.25E+15  &2.94  &0.17  &0.38  &15.3  &1.83E+03  &1.16  &0.84  &5.09E+22  &1.03E-07  &1.06E+04 \\
G028.2726-0.1666 & 280.8799  &-4.2219  & 79.9  & 4.50  &13.5  &0.93  &2.18  &1.06E+16  &4.22  &0.27  &0.56  &10.6  &1.82E+03  &0.17  &0.69  &9.42E+22  &1.13E-07  &1.90E+04 \\
G028.5246-0.2519 & 281.0714  &-4.0368  & 87.7  & 4.69  &13.9  &0.57  &1.35  &6.76E+15  &2.56  &0.20  &0.67  &12.8  &4.08E+02  &0.45  &0.30  &3.44E+22  &1.96E-07  &5.34E+04 \\
G028.5413-0.2371 & 281.0658  &-4.0152  & 86.8  & 4.62  &14.1  &0.74  &1.73  &5.04E+15  &3.27  &0.25  &0.39  &12.9  &1.13E+03  &0.45  &0.65  &6.77E+22  &7.44E-08  &1.43E+04 \\
G029.5561+0.1861 & 281.1531  &-2.9191  & 80.3  & 4.58  &14.1  &0.59  &1.40  &3.86E+15  &2.63  &0.46  &0.17  &12.8  &3.36E+02  &0.45  &0.54  &3.02E+22  &1.28E-07  &7.41E+03 \\
G029.8406-0.0342 & 281.4794  &-2.7665  & 100.2 & 7.67  &18.8  &0.74  &1.73  &1.47E+15  &2.83  &0.29  &0.36  &14.5  &9.26E+02  &1.13  &0.62  &2.96E+22  &4.96E-08  &1.36E+04 \\
G030.0062-0.1192 & 281.6308  &-2.6580  & 99.4  & 6.89  &15.7  &0.65  &1.54  &6.23E+15  &2.75  &0.28  &0.31  &13.7  &7.18E+02  &0.68  &0.58  &2.75E+22  &2.27E-07  &1.26E+04 \\
G030.0556+0.0995 & 281.4585  &-2.5143  & 97.6  & 6.75  &17.2  &0.66  &1.57  &3.78E+15  &2.67  &0.38  &0.23  &14.6  &5.59E+02  &1.17  &0.60  &2.33E+22  &1.63E-07  &9.12E+03 \\
G030.4235-0.2142 & 281.9059  &-2.3301  & 103.8 & 7.23  &20.5  &0.95  &2.23  &3.02E+15  &3.49  &0.54  &0.27  &17.6  &1.00E+03  &3.07  &0.74  &3.41E+22  &8.86E-08  &8.64E+03 \\
G030.5682-0.0258 & 281.8042  &-2.1153  & 88.3  & 4.86  &17.9  &1.01  &2.37  &6.70E+15  &3.98  &0.70  &0.40  &15.0  &5.02E+02  &1.26  &0.43  &3.90E+22  &1.72E-07  &2.23E+04 \\
G030.6574+0.0446 & 281.7823  &-2.0038  & 81.5  & 4.66  &12.6  &0.99  &2.32  &7.45E+15  &4.66  &0.89  &0.22  &14.7  &5.24E+02  &0.87  &0.59  &3.75E+22  &1.99E-07  &9.01E+03 \\
G030.6858-0.0306 & 281.8622  &-2.0129  & 89.7  & 5.00  &22.1  &1.40  &3.29  &1.50E+16  &4.98  &0.66  &0.39  &17.9  &2.23E+03  &3.35  &0.92  &9.07E+22  &1.65E-07  &9.94E+03 \\
G030.7912-0.1173 & 281.9875  &-1.9586  & 94.2  & 5.31  &20.8  &1.72  &4.05  &3.84E+15  &6.32  &3.31  &      &15.9  &3.14E+02  &2.00  &0.43  &2.09E+22  &1.83E-07  &1.34E+04 \\
G030.7941+0.0736 & 281.8189  &-1.8689  & 90.0  & 4.98  &29.5  &0.59  &1.40  &4.13E+15  &1.80  &0.75  &0.11  &15.6  &1.93E+02  &2.23  &0.50  &1.23E+22  &3.37E-07  &5.18E+03 \\
G030.8130-0.0235 & 281.9140  &-1.8964  & 95.2  & 5.77  &20.5  &1.30  &3.05  &1.33E+16  &4.80  &0.32  &1.09  &19.1  &2.48E+03  &4.11  &0.58  &1.36E+23  &9.78E-08  &4.48E+04 \\
G030.8447+0.1775 & 281.7495  &-1.7765  & 96.4  & 6.52  &15.5  &0.78  &1.83  &8.88E+15  &3.29  &0.29  &0.38  &11.0  &1.08E+03  &0.25  &0.64  &5.01E+22  &1.77E-07  &1.40E+04 \\
G030.8523-0.1086 & 282.0077  &-1.9002  & 100.7 & 6.34  &18.5  &0.65  &1.54  &6.70E+15  &2.53  &0.25  &0.26  &17.1  &1.08E+03  &2.82  &0.78  &4.28E+22  &1.57E-07  &7.84E+03 \\
G030.8620+0.0392 & 281.8805  &-1.8242  & 74.7  & 5.13  &10.9  &0.64  &1.52  &4.52E+15  &3.26  &0.33  &0.34  &19.4  &4.49E+02  &2.66  &0.44  &2.99E+22  &1.51E-07  &1.83E+04 \\
G030.8624-0.0394 & 281.9506  &-1.8597  & 94.6  & 5.65  &16.3  &0.56  &1.33  &5.25E+15  &2.32  &0.20  &0.47  &17.9  &5.06E+02  &2.99  &0.40  &3.17E+22  &1.65E-07  &2.76E+04 \\
G031.1496+0.2650 & 281.8108  &-1.4653  & 102.4 & 6.62  &20.9  &0.98  &2.30  &2.35E+14  &3.57  &1.24  &0.10  &20.0  &6.07E+02  &6.65  &0.97  &1.91E+22  &1.23E-08  &2.32E+03 \\
\hline \\
Min              &           &         & 10.5  & 3.12  &10.9  &0.48  &1.14  &2.35E+14  &1.80  &0.13  &0.10  &10.6  &1.93E+02  &0.17  &0.13  &1.23E+22  &1.23E-08  &2.32E+03 \\
Max              &           &         & 114.9 & 14.56 &29.5  &1.72  &4.05  &1.50E+16  &7.36  &3.31  &2.65  &20.3  &4.57E+03  &6.65  &1.12  &1.36E+23  &3.37E-07  &4.89E+05 \\
Median           &           &         & 86.9  & 5.07  &16.3  &0.76  &1.78  &6.08E+15  &3.32  &0.32  &0.39  &14.8  &5.93E+02  &1.15  &0.56  &3.96E+22  &1.47E-07  &1.42E+04 \\
Avg              &           &         & 78.6  & 5.45  &16.8  &0.84  &1.98  &6.59E+15  &3.45  &0.51  &0.51  &15.0  &8.47E+02  &1.70  &0.54  &4.59E+22  &1.50E-07  &3.74E+04 \\
Std              &           &         & 28.9  & 1.85  & 3.5  &0.28  &0.66  &3.33E+15  &1.12  &0.52  &0.46  &2.70  &7.73E+02  &1.56  &0.22  &2.43E+22  &5.56E-08  &7.83E+04